\documentclass{epl}
\newcommand{\f}{\frac}
\newcommand{\pr}{\prime}
\newcommand{\al}{\alpha}
\newcommand{\G}{\Gamma}

\newcommand{\D}{\Delta}

\newcommand{\La}{\Lambda}

\newcommand{\aph}{\alpha}

\newcommand{\Sg}{\Sigma}

\newcommand{\om}{\omega}
\newcommand{\Om}{\Omega}
\newcommand{\ep}{\epsilon}

\def\be{\begin{equation}}
\def\ee{\end{equation}}
\title{Mean field baryon magnetic moments and sumrules}
\shorttitle{Baryon magnetic moments}
\author{M. Bagchi\inst{1} \and S. Daw\inst{1} \and M. Dey\inst{1,2,3}, \and J. Dey\inst{1,2,3}}
\shortauthor{M. Bagchi \etal}
\institute{
  \inst{1} Department of Physics, Presidency College, Kolkata, India \\
  \inst{2} Visitor, ECT, I-38050, Villazzano,  Trento, Italy \\
  \inst{3} CSIR Emeritus Professor }
\begin{document}
\maketitle
\begin{abstract}

New developments have spurred interest in magnetic moments
($\mu$-s) of baryons. The measurement of some of the decuplet
$\mu$-s and the findings of new sumrules from various methods are
partly responsible for this renewed interest. Our model, inspired
by large colour approximation, is a relativistic self consistent
mean field description with a modified Richardson potential and is
used to describe the $\mu$-s and masses of all baryons with up
(u), down (d) and strange (s) quarks. We have also checked the
validity of the Franklin sumrule (referred to as CGSR in the
literature) and sumrules of Luty, March-Russell and White. We
found that our result for sumrules matches better with experiment
than the non-relativistic quark model prediction. We have also
seen that quark magnetic moments depend on the baryon in which
they belong while the naive quark model expects them to be
constant.

\end{abstract}

\section{Introduction}
't Hooft suggested that the inverse of number of colours ($N_c$)
could be used as an expansion coefficient in the otherwise
parameter free QCD \cite{th}. Based on this, Witten \cite{wit1}
suggested mean field (MF) description for baryons - prompting the
use of phenomenological interquark potential tested in meson
sector. Indeed baryon mass ($M$) was calculated at MF level
\cite{ddt} using Richardson potential as an interquark one
\cite{rich}. The potential has confinement and asymptotic freedom
(AF) built in  - with a single scale parameter ($\Lambda \sim
400~MeV$). Although the lattice QCD gives a confinement scale
($\sim 400 ~MeV$) - the AF scale as given by perturbative QCD is ~
$\sim 100 ~MeV$. For computing the hadron properties a $\Lambda
\sim 400~MeV$ is required whereas for high density strange quark
matter (SQM in short) $\Lambda \sim 100~MeV$ \cite{D98}. This is
not surprising as for the SQM the confinement gets screened and a
much reduced $\La $ appropritate for AF part is required. Hence it is
important to separate out the two scales and re-do the hadron
properties -like magnetic moments and masses for the baryons. In
fact, Bagchi et al \cite{mmsj} calculated the masses ($M$) and
magnetic moments $\mu$ of ${\D}^{++}$ and ${\Om}^-$ with a
modified two parameter Richardson potential using relativistic
Hartree-Fock (RHF) method.

The known $\mu$-s cannot be explained by any model exactly and in
addition there are some sumrules which are interesting to look at.
The first one that is much referred is the Franklin sumrule, named
erroneously after Coleman and Glashow \cite{CG}. The latter
authors derived mass sumrules for SU(3) breaking and found
specific $\mu$-s but not the actual sumrule for the octet $\mu$-s
- which was first done by Franklin \cite{fr} and so should carry
his name{\footnote{We are grateful to Jerry Franklin for pointing
this out to us and allowing us to correct the mistake.}}.

Presently, extending the calculation of Bagchi et al \cite{mmsj}
with the modified potential  we find all baryonic (decuplet and
octet) $\mu$-s and check few sumrules to test symmetry
assumptions. We compare with experiment and other calculations
hoping more experimental decuplet $\mu$-s to be soon deduced. Our
values are thus predictions. One can view our calculation in the
spirit of large $N_c$ or consider it as a relativistic MF
calculation with a potential having AF and confinement property
built into it. We compare our results with those obtained from
analytic large $N_c$ formulations which are exact calculations but
assumes static limit as $M$ is infinite for $N_c~=~ \infty$. Our
$M$-s are finite, so the comparison is interesting - displaying
the effect of the wave functions and whatever dynamics it
contains.

\section{Formalism}
One needs to sum all planar gluon exchange diagrams to deduce an
effective interquark potential. Analytic derivation of such a sum
being absent, a potential like Richardson potential is chosen from
meson phenomenology and then tested for baryons and quark stars
\cite{ddt,D98}  \be
 V(r) ~= ~-\f{N_c+1}{2N_c}
  \f{6\pi}{33-2N_f}\left[{\La}^2
r-f({\La} r)\right]\label{eq:richard}
 \ee
$-\f{N_c+1}{2N_c}$ is colour contribution, $N_c$ is number of
colours (= 3), $N_f$ is number of flavours (= 3).
 \be
f(t)~=~1~-~4~\int_1^\infty~\f{dq}q\f{exp(-qt)}{[ln(q^2-1)]^2+\pi^2}
\label{eq:ft} \ee

Richardson calculated, non-relativistically,  the masses of two
heavy mesons, $J/\Psi$ and $\Upsilon$ \cite{rich}. Crater and Van
Alstine \cite{cv} obtained masses of both light and heavy mesons
using a relativistic two body Dirac equation with ${\La}~
=~401~MeV$. Dey $et.~al.$ \cite{ddt} calculated the baryonic
properties like $M$ and $\mu$ of $\Om^-$ with ${\La}~=~400~MeV$
using RHF method. However, when  the potential was  used in
strange star calculation \cite{D98} the required
$\La~\sim~100~MeV$. Strange stars are very compact objects
composed of high density SQM. Debye screening length of the gluon
suppresses the confinement due to the medium effect and  a lower
$\La $ is sufficient.

Separating the confinement and AF scales the modified potential
is given by   \be
 V(r) ~= ~-\f{N_c+1}{2N_c}\f{6\pi}{33-2N_f}\left[\La^{
 \prime~2} ~r-f({\La} r)\right]\label{richtwolam}
 \ee

${\La}^ {\pr} =~350~MeV$ (confinement) and ${\La}=~100~MeV$ (AF)
gave satisfactory results for the hadrons ${\D}^{++}$ and
${\Om}^-$ \cite{mmsj}. Also the strange star calculation improves
\cite{brdd}, - increasing the value of the strong coupling
constant $\alpha_s$ in Debye screening term. This is more
consistent in the framework of the star calculation in view of the
findings of \cite{rdd} where the density dependence of quark
masses is found from $\alpha_s$.

We set out to  apply this potential to find the $\mu$-s of all
other baryons, and check magnetic moment sumrules,  with no more
extra parameter to adjust.

\section{Details of Calculation}

The Quark wave function for quarks in lowest ($1s_{1/2}$) orbital

\begin{eqnarray}
\phi_q(r)=~\left[\f{1}{4\pi}\right]^\f{1}{2}\pmatrix{iG(\vec
r)\chi_m \cr \vec \sigma .\hat{r}F(\vec r) \chi_m} \label{wavefnq}
;~~~~~ q~=u, d, s \end{eqnarray} $\chi_m$ is Pauli spinor, $\vec
\sigma$ is Pauli matrix. The Hamiltonian and corresponding HF
equations are: \be
H~=~{\sum_{i}^3}t_i+\sum_{i<j}{V(r_{ij})}\label{hamil} \ee
\begin{eqnarray}
i=1,2,3;~~ j=1,2,3;~~ t_i~=~{\vec \aph}.{\vec p_i}+{\beta}{m_i}
\end{eqnarray}
\be\left[t_q+{\omega_q(r_1)}\right]\phi_q(r_1)=\epsilon_q\phi_q(r_1)\ee
$t_i$ is the kinetic energy operator for the $i{th}$ quark,
$V(r_{ij})$ is the modified potential, $\epsilon_q$-s are the
single particle energies and $\omega_q$-s are the single particle
potentials. \be \omega_{q1}( r_1)= \int{{\phi_{q2}}^\dagger(r_2)
V(r_{12})\phi_{q2}(r_2){r_2}^2{dr_2}}+\int{{\phi_{q3}}^\dagger(r_3)
V(r_{13})\phi_{q3}(r_3){r_3}^2{dr_3}}\label{eq:wavu} \ee Using
the wave functions given in eqn. [\ref{wavefnq}], we get sets of
coupled differential equations 
\begin{eqnarray}
\nonumber \f{dG_q}{dr}-\left(m_q-{\om}_q+{\ep_q}\right)F_q~=~0 \\
\f{dF_q}{dr}+\left(\f{2}r\right)F_q+\left({\ep_q}-{\om}_q-m_q\right)G_q~=~0
\end{eqnarray}

\noindent From $\ep_q$-s, the energy is obtained from the
following equation

\begin{eqnarray}
\nonumber E=\ep_u+\ep_d+\ep_s-\f{1}{2}\int{{\phi_{u}}^\dagger(r_1)
{{\om}_u}\phi_u(r_1)r_1^2 dr_1}- \\
\f{1}{2}\int{{\phi_{d}}^\dagger( r_2) {{\om}_d}\phi_d(r_2)r_2^2
dr_2} - \f{1}{2}\int{{\phi_{s}}^\dagger( r_3)
{{\om}_s}\phi_s(r_3)r_3^2 dr_3}
\end{eqnarray}

\noindent Here the subtractions are to avoid double counting which
comes when applying variational principle to derive HF equations.

The coupled differential equations are to be solved self
consistently. Procedure is to check the convergence in the energy
value. Moreover, to make the calculation more transparent and
easier we take recourse to expansion of  wave functions in
oscillators and subsequent diagonalisation,  \be
G(r)~=~\sum_n{C_nR_{n0}} \label{eq:g}\ee \be
F(r)~=~\sum_m{D_mR_{m1}}  \label{eq:f}\ee $C$-s and $D$-s are
coefficients. This reduces the differential equations to an
eigenvalue problem. Starting with a trial set of $C$-s and $D$-s, the
solution is found self consistently by diagonalizing the matrix
and putting back the coefficients till convergence is reached. In
general $ R_{nl}(r)$ is  \be
R_{nl}(r)~=~\sqrt{\f{2n!}{{\G}(n+l+\f{3}2)}}r^lexp(-\f{1}2r^2)L_n^{l+\f{1}2}(r^2)
\label{eq:rnl} \ee \noindent $ L_n^{l+\f{1}{2}}(r^2)$ are
Associated Laguerre polynomials. In the calculation, r is replaced
by r/b, b being the oscillator length which may be different for
G(r) and F(r) - $b$ and $b^{\pr}$, respectively.

The centre-of-mass (CM) momentum is not well-defined in RHF
solutions and this entails a spurious contribution from the CM
kinetic energy to the total energy. Since the relative importance
of this effect increases as the number of particles decreases, it
is necessary to correct it for systems formed of few particles.
This has been done here by extending the Peierls-Yoccoz procedure
of nuclear physics. The spurious contribution is denoted by
$T_{CM}$ and the baryon mass M is $E$ - $T_{CM}$. The
Peierls-Yoccoz procedure corrects the energy but  for $\mu$  -
correct boosted wave functions are needed. This is discussed in
the reference \cite{lhw}. Boosting is not attempted in the present
paper since our ultimate aim is to use the procedure for a large
system, a massive star, where these corrections are not relevant.
Therefore, our results on masses of decuplet and octet baryons,
take care of center of mass correction but not for the magnetic
moments. Still, we are able to compare our calculated $\mu -s $
with infinite mass static large $N_c$ models which also do not
involve centre of mass correction{\footnote { Since an oscillator
basis is used for the calculation, it is possible to do standard
CM correction exemplified by Elliott and Skyrme \cite{Elliott}.}}.

Energy differences between decuplet and octet baryons are
obtained using the simplest idea of instanton physics \cite{inst}
where ${\al}$ is the instanton induced potential between a (u,d)
pair and  ${\beta}$ is the same between a (u, s) or (d, s) pair.
\begin{eqnarray}
 E_{N}~=~E_{\D}-3{\al}/2\\
 E_{\La}~=~E_{\La}-{\al}-\beta/2\\
 E_{\Sg}~=~E_{\Sg^*}-3{\beta}/2 \\
 E_{\Xi}~=~E_{\Xi^*}-3{\beta}/2
\end{eqnarray}
\noindent  There are alternative methods also $e.g.$
incorporating colour magnetic interaction energy \cite{ddms}.

The r.m.s. radii $r_{av}$ can also be estimated using the
expression  \be r_{av}~=~\sqrt
{\f{1}{3}{\int_0^{r_{max}}~[\left(G_u(r)^2+F_u(r)^2\right)+\left(G_d(r)^2
+F_d(r)^2\right)+\left(G_s(r)^2+F_s(r)^2\right)]r^4dr}}
\label{eq:rav1}\ee But as the potential is spin independent, the
radii for octet and decuplet members become the same.

In quark model, the magnetic moment associated with a quark is
given by \cite{bhaduri}: \be \mu_q~=~ ~ \f{e_q}2~\int~\left(\vec
r\times \vec j \right)~d^3r \label{eq:muq} \ee \noindent $e_q$ is
the charge of the quark, j is the current associated with the
quark.  $\mu_q$ reduces to : \be
\mu_q[\uparrow(\downarrow)]~=~-(+)e_q ~
\f{2}3~\int_0^\infty~G(r)F(r)r^3dr  \label{eq:muqu} \ee
 Baryonic $\mu$ is found using baryonic wave functions
$\Psi_{spin}\times\Psi_{flavor}$ \cite{close}.

\section{Results}

Subtracting $T_{CM}$ from HF energy $E$, we obtained the baryonic
masses $M$ which are given in table \ref{allbarmy} along with the
experimental values. $T_{CM}$ lies $\sim100~MeV$ for all the
baryons. We have chosen $\al~=~188$ and $\beta~=~103~~MeV$
\cite{inst} which gives the overall best fit. The values of the
oscillator parameters $b$ and $b^{\pr}$ are chosen such that $E$
becomes independent of variation of $b$, $b^{\pr}$. There are no
other free parameters to fit. We have also found that $r_{av}$ is
$\sim~fm$ and it decreases with increasing $M$.

\begin{table}[htbp]
\caption {Charge averaged baryon masses (both decuplet and octet
members) where ${\La}^{\pr}$ is $350~MeV$ and ${\La}$ is
$100~MeV$; using $7\times7$ matrices. The quark masses are
$m_u~,m_d\sim~10~MeV$, $m_s~=~150~ MeV$. We have chosen
$\alpha=~188~MeV$ and $\beta=~103~MeV$ for a good fit of nucleon
and $\Sigma$ comparing to the experimental value.}
\begin{center}
\begin{tabular}{lcccr}\hline
 Baryons &Experimental Mass& Theoretical Mass & &\\
&(MeV)&(MeV)  & $~~~~~b~~~~~$ & $b^{\pr}$ \\ \hline
 ${\D}$'s & 1232& 1251  &0.83 &0.60 \\
& 1383  & & &\\
 ${\Sigma}^*$'s & 1384&1361 &0.83 &0.60 \\
  & 1387 & & &\\
${\Xi}^*$'s &1532 & 1455 &0.77 &0.60 \\
 &1535 &  &&\\
${\Omega}^{-}$ &1672 &1556&0.70 &0.60 \\
  & &938  & & \\
 $N$ &939 &938 &0.83 &0.60 \\
  &1189 & & &\\
 ${\Sigma}$'s&1193 &1188&0.83 &0.60 \\
  &1197 &  & & \\
${\Lambda}^{0}$ &1116  &1098 &0.83 &0.60 \\
 &1315 & & & \\
 ${\Xi}$'s &1321& 1282 &0.77 &0.60\\ \hline
\end{tabular}
\end{center}
\label{allbarmy}
\end{table}

Table \ref{allbarcomp} shows a comparison between our $\mu$-s with
experimental and other theoretical values. The agreement of $M$-s
and $\mu$-s from our result with those from experiments is not too
unreasonable. Results obtained in QCD sum rule (QCDSR) approach
are taken from \cite{monadd} for the decuplet and from
\cite{qcdsroct} for the octet. Results obtained by another large
$N_c$ approximation are taken from \cite{lnc} where the authors
fit the octet and the $\Omega^-$ $\mu$-s to predict the other
decuplet $\mu$-s. Experimental values are taken from \cite{pdg}.
We have also compared our result with lattice \cite{lattice} and
chiral perturbation theory, $\chi pt$ \cite{cpt}. Dai $et.~al$
\cite{dai} fitted the octet $\mu$-s, in their fit A and fit B, by
adjusting 10 parameters and then predicted the unknown $\mu$-s.
For the decuplet the agreement between the calculations and the
three known $\mu$-s are the only guides.

The new experimental value of $\mu_ {\Delta^ {++}}$ agrees better
with our result, QCDSR and lattice than the others.  The Franklin
sumrule  \cite{fr} mentioned in the introduction is as follows :
\begin{eqnarray}
({\mu}_p-{\mu}_n)+({\mu}_{\Sg^-}-{\mu}_{\Sg^+})+({\mu}_{\Xi^0}-{\mu}_{\Xi^-})~=~\Delta_{Franklin}~=~0.
\end{eqnarray}
The value of $\Delta_{Franklin}$ is +0.48 from experimental
$\mu$-s, +0.47 from CDM \cite{cdm} $\mu$-s, +0.14 and +0.30 from
Franklin's recent calculation \cite{fr2} and +0.45 from our
results.

\begin{table}[htbp]
\caption{Comparison of baryon magnetic moments found in different
approaches. } \label{allbarcomp}
\begin{center}
\begin{largetabular}{lccccccccccr}\hline
&${\D}^{++}$ &${\D}^{+}$  &${\D}^{0}$ &${\D}^{-}$
&${\Sigma}^{*+}$ &${\Sigma}^{*0}$ & ${\Sigma}^{*-}$&${\Xi}^{*0}$
&${\Xi}^{*-}$&${\Omega}^{-}$ \\  \hline Ours& 5.77 &2.88 &0.
&-2.86 &2.81 &0.17 & -2.46& +0.30&-2.17 &-1.92 \\
Expt.\cite{pdg} & 6.14 &2.70 &- &-&- &- & -&- &- &-2.02  \\
 QCDSR\cite{monadd}&6.14  &3.02 &0.0 &-3.07 &1.90 &-0.07 &-2.03 &0.80 & -2.71&-2.02 \\
 lattice \cite{lattice}&6.09  &3.05  &0.0 &-3.05 &3.16 &0.33 &-2.5 &0.58 &-2.08 &-1.73 \\
 $\chi pt$ \cite{cpt}&4.0&2.1  &-0.17 &-2.25 &2.0 &-0.07  &-2.2 &0.10 &-2.0 &- \\
$1/{N_c}$ \cite{lnc}&- &3.04 &0.0 &-3.04 &3.35 &+0.32 &-2.79 &0.64 &-2.36 & - \\
 Dai fit A \cite{dai}&5.84&-  &- &- &- &- &- &- &- &-2.08  \\
 Dai fit B \cite{dai}&5.86&- &- &- &- &- &- &- &- &-2.06  \\
&&p & n && ${\Sigma}^{+}$&${\Sigma}^{0}~,~{\Lambda}^{0}$
&${\Sigma}^{-}$  &${\Xi}^{0}$& ${\Xi}^{-}$& \\  Ours &
&2.88 &-1.91 & &2.59  & 0.83, -0.71&-0.92 &-1.45 &-0.62 &   \\
 Expt.\cite{pdg}& &2.79 &-1.91 & &2.46  &$~~~~$ -0.61&-1.16 &-1.25 &-0.65 &  \\
 QCDSR \cite{qcdsroct}& &3.04 &-1.79 & &2.73 & $~~~~$ -0.50&-1.26 &-1.32 &-0.93 &  \\
$CDM$\cite{cdm} &  &2.79 &-2.07 & &2.47&$~~~~$ -0.71 &-1.01 & -1.52&-0.61&   \\
 Dai fit A \cite{dai}& &2.84  &-1.87 & &2.46& &-1.06 &-1.28 &-0.61 &  \\
 Dai fit B \cite{dai}& &2.80  &-1.92 & &2.46 & &-1.23 &-1.26 &-0.63 &  \\ \hline
\end{largetabular}
\end{center}
\end{table}

\begin{table}[htbp]
\caption{Checking of six large $N_c$ analytic sumrules given by
eqn (26) and (30) in Luty, March-Russell and White \cite{luty}
with our results.} \label{luty}
\begin{center}
\begin{tabular}{lcr} \hline
Large $N_c$ analytic relations   & Our results & Experimental\\ \hline
(i) $\mu_{p}+\mu_{n}+\mu_{\Sigma^-}=~0$& +0.05 & -0.28\\
(ii) $\mu_{\Xi^{0}}-2\mu_{\Xi^{-}}=~0$ &-0.21 & +0.05 \\
(iii) $\mu_{\Xi^{-}}+\mu_{\Sigma^{+}}-\mu_{\Sigma^{-}}-\mu_{p}=~0$
& +0.01 & +0.18\\
(iv) $\mu_{\Omega^-}-\mu_{\Xi^{0}}-\mu_{\Xi^{-}}=~0$ &+0.15 &
-0.12\\
(v) $\mu_{\Xi^{0}}+2\mu_{\Sigma^{+}}+2\mu_{\Sigma^{-}}+\mu_{n}=~0$
& -0.02 & -0.56\\  (vi)
$\mu_{\Omega^-}+4\mu_{\Xi^{0}}-3\mu_{\Xi^{-}}+8\mu_{\Sigma^{+}}+5\mu_{\Sigma^{-}}-3\mu_{p}+\mu_{n}=~0$
& -0.29 &-1.486 \\ \hline
\end{tabular}
\end{center}
\end{table}

Taking infinite colour limit and the consequent static infinite
mass one can get some other sumrules among the $\mu$-s
\cite{jm,luty}. Luty $et~ al.$ \cite{luty} coupled the mass
expansion for the s quark to this limit which makes it interesting
to compare with our calculation (table \ref{luty}). Our results
seem to be intermediate to the sumrules and experiment.

With help of equation (\ref{eq:muqu}) we found that $\mu$ for a
particular quark changes in different baryons, whereas quark
$\mu$-s are constant in naive quark model; $\mu_u~=~+1.852$,
$\mu_u~=~-0.972$ and $\mu_s~=~-0.613$ \cite{pdg}. We found that
with increasing M, the magnitude of quark $\mu$ decreases (see table \ref{quark2}).

\begin{table}[htbp]
\caption{Quark magnetic moments in our calculation} \label{quark2}
\begin{center}
\begin{tabular}{lcccccccccr} \hline
&${\D}^{++}$ & ${\D}^{+}$ &${\D}^{0}$ & ${\D}^{-}$ & ${\Sigma}^{*+}$ &${\Sigma}^{*0}$ &${\Sigma}^{*-}$&${\Xi}^{*0}$ &${\Xi}^{*-}$ &${\Omega}^{-}$  \\ \hline
$\mu_u$ &+1.92 &+1.92&+1.91&-&+1.76&-&-&1.64&-&- \\
$\mu_d$ &- &-0.96&-0.95&-0.95&-&-&-0.87&-&-0.82&- \\
$\mu_s$ &-&-&-&-&-0.72&-&-0.72&-0.67&-0.67& -0.64 \\ \hline
\end{tabular}
\end{center}
\end{table}

\section{Conclusions}
The property of  all baryons has been investigated with an
improved Richardson potential in a tree level calculation in the
large $N_c$ spirit and the results agree reasonably with
experiments and other theoretical ones. The calculation is simple
but it gives some of the quark dynamics which is absent in the
exact calculation in infinite $N_c$ (infinite mass) static baryon
model. Accurate determination of other decuplet $\mu$-s will be
helpful for testing models including ours which can be used to
predict the equation of state for SQM.

A different approximate relativistic many body method is perhaps
possible for three quark systems but would not be relevant for us.
We aim to test the validity of MF approach suggested by Witten
\cite{wit1} ``QCD simplifies as $N$ becomes large, and there
exists a systematic expansion in powers of $1/N$. In various ways,
to be discussed later, this expansion is reminiscent of known
phenomenology of hadron physics, indicating that an expansion in
powers of $1/N$ may be a good approximation at $1/N=3$."  and
``The large $N$ limit is, instead, given by a sort of Hartree
approximation. The logic behind this approximation is as follows.
For large $N$ the interaction between any given pair of quarks is
negligible - of order $1/N$. But the total potential experienced
by any one quark is of order one, since any quark interacts with
$N$ other quarks, each with strength $1/N$. Thus, the total
potential experienced by any one quark is of order one, but is a
sum of many small, separately insignificant terms. As in
statistical mechanics, when a quantity is a sum of many
insignificant terms, the fluctuation around the mean value are
very small. Thus, the potential experienced by one quark, apart
from being of order one, can be regarded as a background, c-number
potential-the fluctuations are negligible. To find the ground
state baryon, each quark should be placed in the ground state of
the average potential that it experiences. By symmetry, the
average potential is the same for each quark, so we should place
each quark in the same ground state of the average potential". The
beauty of MF approximation is that it can be used in both 3 quark
system (baryon) and many quark system (quark stars). This has
inspired our work.

\end{document}